\documentclass[manuscript]{aastex}

\begin{document}


\title{Not Color Blind: Using Multi-Band Photometry to Classify Supernovae}


\author{Dovi Poznanski, Avishay Gal-Yam\altaffilmark{1}, Dan Maoz\altaffilmark{2}}
\affil{School of Physics \& Astronomy, Tel-Aviv University, Tel-Aviv
69978, Israel}
\email{(dovip, avishay, dani)@wise.tau.ac.il}

\and

\author{Alexei V. Filippenko, Douglas C. Leonard\altaffilmark{3},
and Thomas Matheson\altaffilmark{4}}
\affil{Department of Astronomy, 601 Campbell Hall, University of
California, Berkeley, CA 94720-3411}
\email{alex@astro.berkeley.edu; leonard@corelli.astro.umass.edu; tmatheson@cfa.harvard.edu}


\altaffiltext{1}{Colton Fellow.}
\altaffiltext{2}{Also at Department of Astronomy, Columbia University, 550 W. 120th
St., New York, NY 10027.}
\altaffiltext{3}{Present address: Department of Astronomy, University of
Massachusetts, Amherst, MA 01003-9305.}
\altaffiltext{4}{Present address: Center for Astrophysics, 60 Garden Street,
Cambridge, MA 02138.}


\begin{abstract} 
Large numbers of supernovae (SNe) have been discovered in recent years, and
many more will be found in the near future. Once discovered, further study of a
SN and its possible use as an astronomical tool (e.g., as a distance estimator)
require knowledge of the SN type. Current classification methods rely almost
solely on the analysis of SN spectra to determine their type. However,
spectroscopy may not be possible or practical when SNe are faint, numerous, or
discovered in archival studies. We present a classification method for SNe
based on the comparison of their observed colors with synthetic ones,
calculated from a large database of multi-epoch optical spectra of nearby
events.  We discuss the capabilities and limitations of this method.  For
example, type Ia SNe at redshifts $z<0.1$ can be distinguished from most other
SN types during the first few weeks of their evolution, based on $V-R$ vs.
$R-I$ colors. Type II-P SNe have distinct (very red) colors at late ($t>
100$~d) stages.  Broadband photometry through standard Johnson-Cousins $UBVRI$
filters can be useful to classify SNe out to $z \approx 0.6$.  The use of Sloan
Digital Sky Survey (SDSS) $ugriz$ filters allows the extension of our
classification method to even higher redshifts ($z = 0.75$), and the use of
infrared bands, to $z = 2.5$. We demonstrate the application of this method to
a recently discovered SN from the SDSS. Finally, we outline the observational
data required to further improve the sensitivity of the method, and discuss
prospects for its use on future SN samples. Community access to the tools
developed is provided by a dedicated website.\footnotemark[5]

\footnotetext[5]{See http://wise-obs.tau.ac.il/$\sim$dovip/typing}

\end{abstract}




\section{Introduction}

The study of supernovae (SNe) has greatly advanced in the last few
years. Intensive and highly automated monitoring of nearby galaxies (e.g., Li
et al. 1996; Treffers et al. 1997; Filippenko et al. 2001; Dimai 2001; Qiu \&
Hu 2001), wide-field, moderately deep surveys (e.g., Reiss et al. 1998; Gal-Yam
\& Maoz 1999, 2002; Hardin et al. 2000; Schaefer 2000), and cosmology-oriented,
deep, high-redshift SN search projects (Perlmutter et al. 1997; Schmidt et
al. 1998) now combine to yield hundreds of new SN discoveries each
year. Ambitious programs that are currently planned or underway [e.g., the
Nearby Supernova Factory -- Aldering et al. 2001; the Supernova/Acceleration
Probe (SNAP) -- Perlmutter et al. 2000; automated SN detections in Sloan
Digital Sky Survey (SDSS) data -- Vanden Berk et al. 2001; Miknaitis et
al. 2001b; see also $\S~4$] promise to increase these numbers by at least an
order of magnitude.

SNe are heterogeneous events, empirically classified into many subtypes, with
the main classification criteria based on spectral properties.  Briefly, SNe of
type II show hydrogen lines in their spectra while SNe of type I do not. Each
of these types is further divided into subtypes, the commonly used ones
including Ia, Ib, and Ic, as well as II-P, II-L, IIn, and IIb. See Filippenko
(1997) for a thorough review and $\S~2$ for more details. It is widely accepted
that SNe~Ia are produced from the thermonuclear disruption of a white dwarf at
or near the Chandrasekhar limit, while all other types of SNe (Ib, Ic, and II)
result from the core collapse of massive stars.

While understanding SNe, their properties, and their underlying physics is of
great interest, SNe are also useful tools in the study of various other
important problems.  SNe~Ia are excellent distance indicators, and their Hubble
diagram has been used to determine the local value of the Hubble constant
(e.g., Parodi et al. 2000, and references therein).  The extension of the
Hubble diagram to higher redshifts ($z \approx 1$) probes the geometry and
matter-energy content of the Universe (e.g., Goobar \& Perlmutter 1995). Two
independent groups using large samples of high-$z$ SNe~Ia presented a strong
case for a current acceleration of the Universe (Riess et al. 1998; Perlmutter
et al.  1999; see Filippenko 2001 for a summary), consistent with a nonzero
cosmological constant, $\Lambda$.  Subsequent work (Riess et al. 2001) based on
a single SN~Ia at $z \approx 1.7$ possibly shows the transition from
matter-dominated deceleration to $\Lambda$-dominated acceleration at $z \approx
1$.  SNe~II-P can also be used as primary distance estimators through the
expanding photosphere method (EPM; Kirshner \& Kwan 1974; Schmidt, Kirshner, \&
Eastman 1992; Schmidt et al. 1994), as was most recently demonstrated by Hamuy
et al. (2001) and Leonard et al. (2002a,b). Leonard et al. (2002a; see also
H\"{o}flich et al. 2001) suggest that distances good to $\sim 30$\% ($1\sigma$)
may be possible for SNe~II-P by simply measuring the mean plateau visual
magnitude, obviating the need for a complete EPM analysis unless a more
accurate distance is desired. Hamuy \& Pinto (2002) refine this technique,
showing that a measurement of the plateau magnitude and the ejecta expansion
velocity potentially yields a considerably smaller uncertainty in the derived
distance.

SN rates as a function of redshift probe the star-formation history of the
Universe, the physical mechanisms leading to SNe~Ia, and the cosmological
parameters (Jorgensen et al. 1997; Sadat et al.  1998; Ruiz-Lapuente \& Canal
1998; Madau, Della Valle, \& Panagia 1998; Yungelson \& Livio 2000). SN rates
are also important for understanding the chemical enrichment and energetics of
the interstellar medium (e.g., Matteucci \& Greggio 1986) and the intracluster
medium (e.g., Brighenti \& Mathews 1998, 2001; Lowenstein 2000; Gal-Yam,
Maoz, \& Sharon 2002).

Once discovered, the study of a particular SN, and its use as a tool for any of
the applications above, is almost always based on spectroscopic verification
and classification. The information extracted from SN spectra usually includes
(but is not limited to) the SN type, redshift, and age (relative to the dates
of explosion or peak brightness).  Spectroscopic followup may not always be
possible or practical.  SNe, especially at high redshift, may be too faint for
spectroscopy, even with the largest, 10-m-class telescopes currently
available. Spectroscopy is also not practical if large numbers (hundreds or
thousands) of SNe are detected within a relatively short time, as is expected
to happen in the case of the SDSS southern strip (Miknaitis et al. 2001b; see
also $\S~4$).  Finally, spectroscopy is impossible for SNe discovered in
archival data (Gal-Yam \& Maoz 2000; Riess et al. 2001; Gal-Yam et al. 2002),
which have long faded by the time they are found.  The discovery of SNe in
archival data is expected to become frequent as high-quality astronomical
databases become larger and more accessible, especially with the development of
projects such as ASTROVIRTEL (\url{http://www.stecf.org/astrovirtel}) and the
National Virtual Observatory (Brunner, Djorgovski, \& Szalay 2001).

The goal of the present work is to facilitate the scientific exploitation of
SNe for which no spectroscopic observations exist.  The obvious alternative for
spectroscopy is multi-color broadband photometry. The potential utility of
such an approach is demonstrated, in principle, by the use of the ``photometric
redshift'' method to infer the redshift and type of galaxies and quasars that
are too faint or too numerous to observe spectroscopically (e.g., Weymann et
al. 1999; Richards et al. 2001).  Applying a similar method to SNe is not
straightforward since, unlike galaxies, the spectra of SNe vary strongly with
type, redshift, and time. While photometric approaches to the study of faint
SNe have been discussed before (Dahl\'en \& Fransson 1999; Sullivan et
al. 2000; Riess et al. 2001), no general treatment has been presented thus far.

In a recent paper, Dahl\'en \& Goobar (2002, hereafter DG2002) tackle the issue
of SN classification for the case of deep, cosmology-oriented, high-$z$ SN
searches. Their treatment relies heavily on the particular observational setup
used by these programs --- i.e., the comparison of images obtained $\sim 3$--4
weeks apart, designed to detect SNe before, or at, peak brightness. As these
authors show, this setup is strongly biased toward the discovery of SNe~Ia. The
main theme of DG2002 is the selection of the high-$z$ SNe~Ia from an observed
sample, using photometry of the host galaxies and the SNe themselves. For the
particular observational setup they consider, DG2002 also wish to minimize the
contamination of the sample by non-Ia SNe. To do this, they use the models of
Dahl\'en \& Fransson (1999) and demonstrate their ability to reject most non-Ia
SNe based on their colors and lower brightness. However, the models used for the
population of non-Ia SNe are based on many parameters and assumptions, such as
the ultraviolet (UV) spectra, peak brightness, light curves, and relative
fractions of non-Ia SNe, that are not precisely known even in the local
universe, let alone at high redshift. While DG2002 seek only to identify
high-$z$ SNe~Ia detected by a particular observational setup, we present general
methods that apply to all the major SN subtypes and can be used for a wide range
of SN surveys, including searches that are similar to those described by DG2002.
   
Classification of SNe using broadband colors is such a complex problem that, as
we show below, a full solution (i.e., deriving SN type, redshift, and age from a
few broadband colors) is probably impossible. However we will argue here that
the problem may be simplified.  Type, redshift, and age need not all be
determined simultaneously. Most SNe are associated with host galaxies that are
readily detectable. One can therefore infer the redshift of the SN from that of
the host, derived either from spectroscopy (that may be obtained long after the
SN has faded) or using a photometric redshift. In fact, for SNe that are
detected in well-studied parts of the sky, the redshift information often
already exists. Gal-Yam et al. (2002) recently demonstrated this with SNe they
discovered in archival {\it Hubble Space Telescope (HST)} images of well-studied
galaxy-cluster fields. Five of the six apparent SNe they found (and all of those
clearly associated with a host) had redshift information in the literature. For
the majority of SNe, the redshift can thus be obtained from observations of the
host, and treated as a known parameter.

In the absence of a spectrum, the SN age may sometimes be revealed by measuring
the photometric light curve of the event. Admittedly, such followup may still
pose a problem for faint or very numerous events, and is generally not possible
for archival SNe (see Riess et al. 2001 for an exception). However, for many of
the scientific uses of SNe (e.g., the derivation of SN rates), the age of each
event is immaterial.

Determination of the type of a SN is crucial for most of the applications
discussed above.  The SN type can be securely determined only from observations
of the active SN itself. While some information can be gleaned from the host
(e.g., elliptical galaxies are known to contain only SNe~Ia), this information
is limited and such reasoning may not hold at high redshifts.  We therefore
concentrate our efforts on using multi-color broadband photometry to classify
SNe.




\section{Method}

Since we want to constrain the type of a SN with an arbitrary (but known)
redshift, it is not simple to utilize broadband photometry; K-corrections for
SNe of all types, and at all ages, are currently unknown. Instead, we have
compiled a large spectral database of nearby, well-observed SNe. These spectra
are used to calculate synthetic broadband colors for various SN types through a
given filter set at a given redshift.  The temporal coverage of our compilation
allows us to draw paths in color space which show the time evolution of each SN
type.  Inspecting the resulting diagrams, one can then look for regions which
are either populated by a single type of SN, or that are avoided by various SN
types.  The observed photometric colors of a candidate SN with a known redshift
can then be plotted on the relevant diagrams, and constraints on its type and
age may be drawn.  As we show in $\S~3$, the type of a SN can sometimes be
uniquely determined. When this is not the case, the type may still be deduced
by supplementing the color information with other available data on the SN,
such as constraints on its brightness, and information (even if very limited)
on its variability (e.g., whether its flux is rising or declining).

\subsection{Supernova Subtypes}

Our SN classification method is based on colors, determined by the spectral
energy distribution (SED) of each event. By definition, such a method can
differentiate only between SN subtypes with unique spectral
characteristics. Hypothetically, two SNe with very different photometric
properties (e.g., light curves, peak magnitudes), but having similar SEDs at
all ages, would not be distinguished by our method. The SN classification
scheme we adopt needs, therefore, to rely only on spectral
properties. Fortunately, the common classification of SNe (see Filippenko 1997
for a review) is mostly based on their spectra, with one notable exception
which we discuss below.

It is customary to divide the SN population into two main subgroups: type II
SNe, which show prominent hydrogen lines in their spectra, and type I SNe,
which do not (Minkowski 1941). Type I SNe are further divided into type Ia
whose spectra are characterized by a deep absorption trough around $6150$~\AA,
attributed to blueshifted Si~II $\lambda\lambda$6347, 6371 lines, type Ib that
show prominent He~I lines, and type Ic that lack both Si~II and He~I (e.g.,
Matheson et al. 2001, and references therein).

Type II SNe are divided (e.g., Barbon, Ciatti, \& Rosino 1979; Doggett \&
Branch 1985) into two subclasses according to their light-curve shape: SNe~II-P
show a pronounced plateau during the first $\sim 100$~d of their evolution, and
SNe~II-L decline in a linear fashion in a magnitude vs. time plot, similar to
SNe~I.  Further studies (e.g., Filippenko 1997) show that SNe~II-P have
characteristic spectra, with the H$\alpha$ line showing a P-Cygni profile. On
the other hand, SNe~II-L have not been well characterized spectroscopically,
and the best spectroscopically studied events (SNe 1979C and 1980K) are
considered to be photometrically peculiar (``overluminous''; Miller \& Branch
1990) relative to other SNe II-L.  During the 1990s, the focus in studies of
SNe~II shifted from photometry to spectroscopy. Two new subtypes have
emerged. SNe~IIn show narrow H$\alpha$ emission (e.g., Schlegel 1990;
Filippenko 1997, and references therein), and SNe~IIb are transition objects
that first appear as relatively normal SNe~II, but evolve to resemble SNe~Ib
(Filippenko 1988; Filippenko, Matheson, \& Ho 1993; Matheson et al.  2001, and
references therein).  For our study, only SN subtypes with well-defined
spectral properties are meaningful. We therefore consider henceforth only SNe
II-P, IIn, and IIb as subtypes of the SN~II population. SNe~IIn and IIb are
probably a large fraction of the SN population that was previously classified
as type II-L based only on their light-curve shape (approximately half of the
SN~II population, e.g., Cappellaro et al. 1997).  Reviewing the latest spectral
databases available to us, we estimate that most of the SNe~II discovered by
current surveys belong to the spectroscopically defined subtypes II-P, IIn, and
IIb (see below).

SNe are a veritable zoo, with many peculiar and even unique events, from
luminous hypernovae (e.g., SN 1998bw --- Galama et al. 1998; Patat et al. 2001;
SN 1997cy --- Germany et al. 2000; Turatto et al. 2000; SN 1997ef --- Iwamoto
et al. 2000; Matheson et al. 2001) to faint SN 1987A-like events (e.g., Arnett
et al. 1989, and references therein), and SN ``impostors'' (e.g., Filippenko et
al. 1995a; Van Dyk et al. 2000, and references therein).  Currently, we limit
our discussion to SN types that are not extremely rare, and are well-defined
and well-characterized spectroscopically.  Assuming that the SNe reported in
the IAU Circulars are representative of the SN population that is discovered by
current programs, we estimate that, collectively, the well-defined subtypes we
consider (Ia, Ib, Ic, II-P, IIn, and IIb) constitute at least 83\%, and
probably more, of the entire population.

For instance, we consider all the SNe that were discovered during the year 2000
and whose type was either reported in IAU Circulars, or alternatively could be
constrained by data available to us. Among 110 such events, there were 54 type
Ia, 5 type Ib, 5 type Ic, 8 type IIn, one type IIb, two peculiar SNe~Ia, one
peculiar SN~Ib, and 34 events which were reported just as ``type II.'' Using
available spectra or data reported in the IAU Circulars, we learn that of the
latter 34 events, 18 show a P-Cygni profile in H$\alpha$, typical of SNe~II-P,
13 events may or may not belong to the II-P, IIn, or IIb subtypes, and 3 events
are definitely not II-P, IIn, or IIb events. Hence 91 events (83\%) belong to
our spectroscopically defined SN sample (types Ia, Ib, Ic, II-P, IIn, and IIb),
13 events (12\%) may or may not be of these types, and just 6 events (5\%) are
known to be of types not included in our database. It is therefore likely that
analysis based on our database will be relevant to the large majority of SNe
discovered by current programs.  This is probably also true for future programs
that will use similar search methods. We are aware that a large population of
intrinsically faint SNe (e.g., SN 1987A-like) may be underrepresented in
current SN statistics.  Analysis of future surveys that will be sensitive to
such events may require their inclusion in the spectral database.

\subsection{Spectral Database}

The core of our classification algorithm is the spectral database: a
compilation of optical spectra of nearby SNe. Most of the these spectra were
obtained with the Shane 3-m reflector at Lick Observatory. The typical
resolution is better than $15$~\AA\ (full width at half maximum), and the
majority of the data used cover the range 3200--10000~\AA. Some spectra with
narrower coverage (approximately 4000--8000~\AA) are also used.  All spectra
were either obtained at the parallactic angle (Filippenko 1982), or calibrated
by simultaneous photometry, to ensure that spectral shape distortions due to
wavelength-dependent atmospheric refraction are not significant. Telluric lines
were removed, generally through division by the spectrum of a featureless
star. 
According
to the Galactic coordinates of the observed SN, the spectra were corrected for
Galactic reddening, using the maps of Schlegel, Finkbeiner, \& Davis (1998) and
the extinction curve of Cardelli, Clayton, \& Mathis (1989).

The working edition of the database was constructed in the following manner.
For each of the well-defined SN subtypes mentioned above, we have selected a
prototype. The main criterion was the availability of spectra with wide
wavelength coverage and multiple epochs. In addition, a few complementary
spectra of other events were included in order to verify consistency and fill
gaps in the temporal evolution, when needed. Table 1 lists the SNe whose
spectra constitute our database, with the prototypical event for each subtype
listed first.

We have specifically avoided events that were known to either suffer
considerable extinction, or to be otherwise peculiar.  For SNe~Ia, peculiar
(overluminous, SN 1991T-like, or underluminous, SN 1991bg-like) events are
rather well characterized, and may be quite common, at least in low-redshift
environments (Li et al. 2001a). Our main database includes only spectra of
normal SNe~Ia, but we examine the effects of peculiar SNe~Ia in $\S~3.4$ below.

All spectra were de-redshifted according to the host-galaxy redshift taken from
the NED database. After redshifting the spectra again to a chosen redshift (see
below), the telluric features were re-applied before deriving synthetic colors.

\subsection{Color-Color Diagrams}

Using our spectral database, we calculate the synthetic colors of SNe at a
chosen redshift through the Johnson-Cousins $UBVRI$ (Johnson 1965; Cousins
1976; see Moro \& Munari 2000 for details), Bessell $JHK$ (Bessell \& Brett
1988), and SDSS $ugriz$ (Fukugita et al. 1996; Stoughton et al. 2002) filter
systems. If a filter's bandpass is not fully covered by a SN spectrum, the flux
in the missing spectral region is extrapolated linearly using the median value
of the spectrum. Each calculated magnitude that includes such an extrapolation
is assigned an error equal to the amount of flux in the extrapolated part of
the spectrum, thus providing a conservative error estimate. The resulting
synthetic photometry is then displayed on color-color diagrams. Note that in
all the plots we present, the ordinate and abscissa are composed of colors
consisting of adjacent bands (e.g., $U-B$ vs. $B-V$, $B-V$ vs. $V-R$
etc.). Since most of the spectra in our database have a wide wavelength range,
error bars will hence generally appear only on one axis.

In the following section we present such diagrams, discuss their use in the
classification of SNe discovered by various programs, and derive a few general
rules. Note that the choice of colors that gives the best SN type
differentiation, apart from the obvious dependence on the filter system used,
also depends on redshift. For optimal results, for each SN (having a known
redshift and some observed colors) one needs to search for color-color diagrams
that give the maximum information content. Obvious space limitations allow us
to present below just a few representative examples of such plots. Similar
plots for arbitrary values of $z$ may be obtained from our website
(\url{http://wise-obs.tau.ac.il/$\sim$dovip/typing}).

\section{Results} 

\subsection{Comparison with Available Photometry}

In Figure 1, we compare our data on SNe~Ia to template colors of normal SNe~Ia
compiled by Leibundgut (1988).  One can see that our calculated colors follow
the expected path in color space, with a small scatter ($\sim 0.2$ mag),
consistent with the previously measured dispersion in SNe~Ia colors (e.g.,
Tripp 1998; Phillips et al.  1999, and references therein).  Since our spectral
database includes spectra of SN 1994D, which was somewhat bluer at early times
than average SNe~Ia (Richmond et al. 1995), our treatment covers an even wider
variety in the colors of early SNe~Ia. From this plot, we can estimate the
expected width of color-color paths for normal SNe~Ia to be $\sim 0.2$--0.3
mag.

Note that SNe~Ia follow a complex curve in color space, with a distinct
turn-around point $\sim 30$~d after peak brightness.  The underlying spectral
evolution is nontrivial.  Like most SNe, the continuum slope of early SNe~Ia is
very blue, and becomes redder as the SN grows older.  This process is dominant
during the first weeks of SN~Ia evolution, but is later countered by the
emergence of strong emission lines in the blue part of the spectrum (restframe
4000--5500~\AA). The line-dominated spectrum results in unique colors for
late-time ($t > 100$~d) SNe~Ia (e.g., Figure 3, below).

In Figure 2 our results for SNe~II-P are plotted against the observed
photometry of the recent, well-observed SN 1999em.  It can be clearly seen
that the results agree perfectly up to around 70~d past maximum brightness,
where a bifurcation appears in our calculated color path. Our spectral data
for SNe~II-P is based on three different events (SNe 1992H, 1999em, and
2001X). The three points that have considerably lower $U-B$ values (at 110,
127, and 165~d) have all been calculated from the spectra of SN 2001X, which
up to that age was consistent with the other SNe~II-P used. This demonstrates
the well-known variety of late-type spectra of SNe~II (Filippenko 1997). In
subsequent plots, we retain both branches of ``old'' SNe~II-P (with the one
derived from spectra of SN 2001X marked, when significantly different, in a
darker shade), and treat the color space bounded by them as a possible
location for such SNe.
 
Contrary to SNe~Ia, the color path of SNe~II-P is quite smooth. The underlying
spectral evolution is driven mostly by the change in the continuum slope, from
very blue at early ages, growing redder with time (e.g., Filippenko
1997). Late-time ($t > 100$~d) spectra become increasingly dominated by strong
and broad emission lines, especially H$\alpha$, that give SNe~II-P their
distinct late-time colors (see, e.g., Fig. 3, below).

\subsection{Zero-Redshift Diagrams}

In Figure 3 (top panel), we show the color paths of all types of SNe at zero redshift in
$U-B$ vs. $B-V$. One sees that many SNe before maximum light are blue, with
$B-V<0.3$ mag. While most young SNe~II have $U-B \approx -0.6$ mag, young
SNe~Ia have bluer $B-V$ colors ($\sim-0.1$ mag) but redder $U-B$ colors, around
$U-B = -0.4$ mag. It is also evident that old ($>100$~d) SNe~Ia have unique
colors ($B-V < 0.4$ mag, $U-B > 0.5$ mag), as do some of the SNe~II-P (ages
19--100~d) which have the reddest colors of all SNe. The arrow shows the
reddening effect corresponding to $A_{V}=1$ mag of extinction, assuming the
Galactic reddening curve of Cardelli et al.  (1989). One can see that the
unique colors of young ($t < 12$~d) and very old ($t > 100$~d) SNe~Ia cannot be
masked even by significant reddening in their host. Furthermore, we note that
high extinction values are rare for SNe~Ia. For example, only 3 among 49 SNe~Ia
($\sim 6 \%$) studied by the CfA and Cal\'{a}n/Tololo programs, and presented
in Riess et al.  (1999), have $A_{V} > 1$ mag. Conversely, because reddening
works only in one direction, some candidates with appropriate observed colors
can be uniquely determined to be SNe~Ia.

However, the center of the diagram is populated by SNe of all types. For
example, a zero-redshift SN, of unknown type, with colors of $U-B = 0.2$ mag,
$B-V = 0.5$ mag, would be impossible to classify based on these colors
only. But the middle panel of Figure 3 shows that if one examines an additional
color, such as $V-R$, this degeneracy can be partly lifted. For the same $B-V =
0.5$~mag, there is a spread in $V-R$ that can shed some light on the SN type. A
value of $V-R \approx 0$ mag would favor a type Ia classification, while a
positive value would suggest one of the core-collapse types.  Addition of an
$R-I$ measurement (Fig. 3, bottom panel) would narrow the options even further,
since in this color, at this redshift, different types have distinct values,
from the bluest SNe~Ia to the reddest types (Ib, Ic, and II-P).  Looking at the
reddening vectors in the middle and bottom panels of Figure 3, we again note
that only heavy extinction ($A_{V}>1$ mag) can move SNe~Ia into color regions
populated by core-collapse SNe.

Before we proceed, we briefly comment on the effects of host-galaxy
contamination. Since our approach assumes the redshift of a given SN is a known
parameter, and this is commonly available through the observation of the host
galaxy, some data on the galaxy should exist. It should therefore generally be
possible to subtract the galaxy's light from the SN photometry. To study the
effect of non-subtraction of host-galaxy light, or imperfect subtraction, we
parameterize the amount of the contaminating galactic flux, at a given epoch,
by the ratio $C = F_{galaxy}/F_{SN},$ where $F_{galaxy}$ and $F_{SN}$ are the
mean flux of the host galaxy and SN spectra, respectively, calculated over the
wavelength range of the spectrum of the SN.  We consider contamination as
negligible, when it is not likely to make one type of SN look like another.
First, we find that for all types of SNe and hosts, when $C < 0.1$ the effect
of contamination is indeed negligible. For young SNe~Ia, the worst-case
scenario would be the contamination by a red elliptical host that could make
them look like redder core-collapse SNe. When observed in restframe blue colors
(e.g., $UBV$), this effect becomes significant only when $C \geq 1$, but if red
bands (e.g., $VRI$) are used, contaminated SNe~Ia attain colors similar to
those of the core-collapse population already at $C = 0.5$. The opposite
effect (i.e., of bluer host galaxies on red core-collapse SNe) is much weaker,
so contamination by spiral hosts is unlikely to mix core-collapse SNe with
young SNe~Ia. Only heavy contamination ($C > 1$) by a very blue, star-forming
host galaxy can make a red SN~II-P appear similar to a bluer SN~Ia on a $UBV$
color-color diagram. In redder bands (e.g., $VRI$), even this extreme case
causes negligible shifts in color. We conclude that if the host galaxy light
has been subtracted, at least roughly, from the measured SN photometry,
residual contamination has no significant implication. A large contribution
from an underlying host (e.g., with a flux similar to that of the SN itself)
that has not been removed may mask blue SNe~Ia as core-collapse events, but is
unlikely to make red core-collapse events appear as SNe~Ia. In the discussion
below, we therefore neglect contamination of the SN photometry by light from
the underlying host galaxy.
 
\subsection{High-Redshift Diagrams}

With increasing redshift, the available spectral information shifts to
longer-wavelength filters. The top panel of Figure 4 demonstrates how well
SNe~Ia can be differentiated from other types at $z = 0.1$ on a $V-R$ vs. $R-I$
diagram. On the $R-I$ axis the only SNe with negative values are SNe~Ia. They
are also the only type with $V-R$ color greater than 0.6 mag while $R-I$ is
smaller than 0.3 mag. During most of the temporal evolution of SNe~Ia, their
distinction from other types is practically unaffected by host galaxy dust
reddening, as can be seen from the vector plotted (the reddening is calculated
in the SN rest frame).

The middle panel of Figure 4 shows a similar diagram computed at $z=0.5$.
Classification is obviously more difficult. Still, SNe~II-P (including ``SN
2001X-like,'' see $\S~3.1$) at ages above 100~d have $R-I$ values higher than
SNe of other types (at all ages). Close to peak flux, SNe~Ia have $V-R$ colors
of a few tenths of a magnitude bluer than other types, and again, this
distinction is not sensitive to reddening. Even where all types seem to mingle,
at $R-I \approx 0.7$ mag and $V-R \approx 1.1$ mag, SNe~Ia and II-P have $V-R$
values that are 0.2 mag higher than other types. Should one of these types be
rendered unlikely by other data, such as a SN that is too bright to be a
SN~II-P, the classification can become unique or nearly so.  For example,
SNe~II-P can be unambiguously identified by their light curves; no other SN
type has a roughly constant $VRI$ brightness over such a long period. In
principle, two $V$-band (restframe) observations separated by, say, 30~d can
unambiguously identify a SN~II-P.

At redshifts higher than 0.6, the $V$ band samples the restframe UV, not
covered by our spectral database. Thus, only one color ($R-I$) remains in the
Johnson-Cousins system.  The SDSS filter system has 3 filters ($r$, $i$, and
$z$) in the approximate range covered by the $R$ and $I$ filters, but extending
farther into the red by $\sim 1000$~\AA. Using this system, we can extend our
analysis to higher redshifts.  This is demonstrated in the bottom panel of
Figure 4 for SNe at $z = 0.75$. Again, late-time SNe~II-P are significantly
redder than other types ($r-i > 1.5$ mag and $i-z > 1$ mag), while early-time
SNe~Ia are significantly bluer ($r-i < 1$ mag and $i-z < -0.2$ mag), and remain
so even in the presence of significant reddening by dust.  At later times (over
two months), SNe~Ia escape from the vicinity of other types and reach $r-i$
values above 1.2 mag. SNe~IIn at this redshift appear to be isolated from other
types.

As we move even farther up the redshift scale the use of infrared (IR) filters
is required. Between $z = 0.8$ and $1.3$, classification would require a
combination of optical and IR photometry. Figure 5 (top panel) shows that at
$z=1$, SNe~Ia have distinct blue colors ($I-J < 2$ mag and $J-H < 0.1$ mag) at
almost all ages, SNe~II-P develop very red $I-J$ colors ($> 2.8$ mag) two
months after peak, while SNe~Ib and Ic are the only ones that have $J-H$ colors
that are redder than $0.7$ mag. As before, these distinctions are hardly
affected by extinction. Between $z = 1.3$ and 2.5 our spectra shift into the
region covered by the Bessell $JHK$ filters. In the bottom panel of Figure 5
($z = 1.5$), one can see that SNe~Ia have generally lower $J-H$ colors ($< 0.5$
mag), except between ages of one to two months, shared perhaps by very early
type IIn, II-P, and Ic events. SNe~Ib and Ic are the only objects to have both
$J-H$ and $H-K$ above unity.

The use of synthetic colors, calculated from spectra of local SNe, to classify
high-$z$ events, assumes that SN spectra do not evolve strongly with
redshift. The similarity, to a first approximation, between the spectra of
distant SNe~Ia and their local counterparts has been demonstrated (Riess et
al. 1998; Coil et al. 2000). However, for all other types of SNe, high-quality
spectra of distant ($z > 0.1$) events are largely unavailable.  Since we use
synthetic broadband colors, only significant evolution in either the spectral
slope or strong emission or absorption features will influence our
classification method. Nevertheless, this is a caveat that must be addressed
once spectra of distant core-collapse SNe become available.

\subsection{Peculiar Type Ia SNe}

We now examine the colors of peculiar SNe~Ia and their implication for our
classification method.  Following the same approach used earlier, we have
compiled spectra of the most extreme varieties of peculiar SNe~Ia: overluminous
(SN 1991T-like; e.g., Filippenko et al. 1992a) and underluminous (SN
1991bg-like; e.g., Filippenko et al. 1992b) objects, as well as spectra of SN
2000cx, a unique SN~Ia that may be considered a subtype of its own (Li et
al. 2001b). Table 2 provides observational details for these objects.

Inspecting Figure 6 (top panel), which shows the $U-B$ vs. $B-V$ colors of the
various SN~Ia subtypes at zero redshift, we see that, although the paths of
peculiar SNe~Ia in color space are different from those of ``normal'' SNe~Ia,
they populate the same general region.  On the other hand, in the middle panel
of Figure 6, showing $V-R$ vs.  $R-I$ colors at zero redshift, one can see that
underluminous SNe~Ia have significantly larger $R-I$ values (by 0.2--0.6 mag)
than those of all other SNe~Ia, and are therefore liable to mix with the redder
core-collapse population.  This is illustrated in the bottom panel of the same
figure, where indeed, underluminous SNe~Ia at early ages (prior to $\sim 60$~d)
cannot be distinguished from core-collapse SNe. Later in their evolution, these
events acquire unique colors ($V-R < 0.2$ mag, $R-I > 0.3$ mag) that are
different from all other types of SNe.
 
As SN 1991T-like and SN 2000cx-like events are almost always bluer than
``normal'' SNe~Ia, they are generally easier to distinguish from core-collapse
SNe, and pose no special problem in our analysis. SNe~Ia of the underluminous,
redder, SN 1991bg-like variety are hard to distinguish from core-collapse SNe
at early ages, and may be lost in color-based classification. This problem is
quite limited at low redshift, where underluminous SNe~Ia are rare ($\sim 16$\%
of the SN~Ia population; Li et al. 2001a), and may be totally negligible at
higher redshifts where, empirically, SN 1991bg-like events have not been found
at all (Li et al. 2001a).

\subsection{Applications}

\subsubsection{Distinguishing Between Type Ia and Ic SNe at High Redshift}

High-redshift SN search programs that focus on finding SNe~Ia for use as
distance indicators encounter the problem of sample contamination by other SN
types, especially a luminous variety of SNe~Ic (Riess et al. 1998; Clocchiatti
et al. 2000).  The problem is two-fold. First, unwanted non-Ia SNe take up
precious telescope resources for followup spectroscopy.  Second, when only low
signal-to-noise ratio spectra of faint, high-$z$ events are available, luminous
SNe~Ic may masquerade as SNe~Ia. The latter problem is acute at high $z$, since
at these redshifts the SN~Ia hallmark (the Si~II $\lambda6355$ trough) is
shifted out of the optical range, and SNe~Ic spectra lack the telltale H or He
lines that distinguish SNe~II and SNe~Ib, respectively.  Spectra with high
signal-to-noise ratios can provide definitive classifications (Coil et
al. 2000), but these are time-consuming to obtain.

>From Figure 7, one sees that color information may help, at least partially, to
alleviate this problem. We confirm and quantify the well known ``rule of
thumb'' (e.g., Riess et al. 2001): SNe~Ic are red compared to SNe~Ia at a
similar redshift.  One can see that, for SNe at ages that are relevant for
high-$z$ search programs, designed to discover SNe near peak brightness, SNe~Ia
are typically 0.5 mag bluer in $r-i$ than SNe~Ic, well above the typical
effects of reddening. While it is true that rising (pre-maximum) SNe~Ic have
colors similar to those of older ($\sim 2$ weeks past maximum) SNe~Ia, even the
most basic variability information (e.g., whether the object is rising or
declining) breaks this degeneracy. Figure 7 also illustrates the fact that most
of the information at this redshift can be obtained from the $r-i$ color alone,
and the inclusion of $z$-band information does not appear to be cost-effective.

\subsubsection{Classifying One of the First SDSS SNe}

SN 2001fg was the third SN reported from SDSS data. This event was discovered
on 15 October 2001 UT by Vanden Berk et al. (2001) with $g = 19.20$, $r =
17.87$, and $i = 18.17$ mag. Inspecting the color-color diagram calculated for
the appropriate redshift ($z = 0.0311$, see below), one can see (Fig. 8) that
the type and approximate age of the SN candidate can be deduced.  The diagram
clearly indicates this is a SN~Ia, around one month old.  Followup spectra by
Filippenko \& Chornock (2001) using the Keck II 10-m telescope reveal that the
object is indeed a SN~Ia, at $z = 0.0311$. The spectrum is similar to those of
SNe~Ia about two months past maximum brightness.  This age, at the time of the
spectral observation (18 November UT), implies an age around one month past
maximum brightness at discovery, confirming our diagnosis. A fully ``blind''
application would have, of course, required an independent photometric or
spectroscopic redshift for the SN host galaxy.

\section{Future Prospects and Conclusions}

The method presented here can become significantly more powerful when more
observational data become available. Faint, high-$z$ SNe are usually observed
at optical wavelengths (i.e., $V$, $R$, or $I$ bands). In order to use these
bands for classification, knowledge of the restframe UV spectrum is
required. Currently, UV spectra are available for only a handful of objects,
while methods like those discussed here require multi-epoch coverage of several
objects for each SN subtype.

Further characteristics of the various SN types, most notably the absolute peak
magnitudes and their scatter for every SN type, the typical light curves, and
the relative rates of the different types, would also be useful. With such data
in hand, one could use the limiting magnitude of a survey to calculate, for
each SN type, at what epochs it is bright enough to be observed, thus removing
some degeneracy in the classification.  A further enhancement could be
introduced to deal with the regions in color space that are populated by
several different SN types. Using the typical light curves and known peak
magnitudes for all types, and the relative rates, one could calculate the
amount of time a SN of a certain type spends within a region and deduce the
likelihood that an observed event belongs to a particular type. As already
mentioned in $\S~1$, work along these lines has been presented by DG2002,
who replace missing observational data with rough estimates or models.

SNe are generally discovered by their variability. However, the tools developed
in this paper may enable one to detect SNe by their colors alone.  Indeed, such
methods have recently been discussed for the detection of another kind of
transient phenomenon --- gamma-ray burst afterglows (Rhoads 2001). Multi-color
wide-field surveys (e.g., Becker et al. 2001; Jannuzi et al. 2001; SDSS ---
York et al. 2000) that are either planned or already underway may be used to
discover large numbers of SNe using similar methods.  To illustrate this point,
we plot in Figure 9 the locations of SNe of all types at $z = 0.2$, together
with a polygon within which fall the vast majority of stars, asteroids,
cataclysmic variables, Cepheids, and quasars, as discussed by Krisciunas,
Margon, \& Szkody (1998).  Except for late-time SNe~Ic, all SNe have lower
$i-z$ values and higher $r-i$ values than these objects. Similar results are
obtained for the SN population at $z =$ 0--0.3, the redshifts of SNe probed by
the SDSS.  This demonstrates the feasibility of color selection of SNe, a
method which may emerge as an invaluable resource when applied to upcoming,
large data sets.

An exciting application of the methods presented in this paper would be the
study of SNe discovered in the SDSS southern strip.  During the months of
September to November, when the primary target of the SDSS, the north Galactic
pole, is not visible, the SDSS telescope observes parts of the sky near the
south Galactic pole. The so called ``south equatorial strip'' is a patch of sky
that will be imaged repeatedly during the 5-year planned operations of the
survey. According to the SDSS website
(\url{http://www.sdss.org/documents/5yearbaseline.pdf}), this 270 square degree
area will be imaged $\sim 19$ times during the 5-year period, or about 4 times
each fall, on average. Assuming that this sampling enables the discovery of all
SNe that occur during the $\sim 3$ months this strip is imaged each year, for
five years, the total surveyed area is $270 \times 3 \times 5$ deg$^2$-months,
or 337.5 deg$^2$-years. Pain et al. (1996) have measured a SN~Ia rate of 34 per
year per square degree in the magnitude range $21.3 \le R \le 22.3$, and the
limiting magnitude of the SDSS images in this band is given as $r = 22.5$
mag. We would then expect some 2300 SNe~Ia each year or a total of over 11,000
SNe~Ia in this strip. The numbers of SNe of all types (not just Ia) will
undoubtedly be even larger. Miknaitis et al. (2001b) report that a systematic
search for SNe in these data is underway, and the first SNe have already been
reported (Miknaitis et al. 2001a; Rest, Miceli, \& Covarrubias 2001; Miknaitis
\& Krisciunas 2001).

This huge expected dataset, which may contain an order of magnitude more events
than all previously known SNe, all with five-band photometry and possibly some
variability information, offers an unprecedented resource for SN studies. As an
example, SN rates, based on thousands of events, can be computed as a function
of SN type, host-galaxy morphology, and redshift.  However, it is clear that
obtaining spectra for thousands of faint SNe each fall is not
feasible. Photometric classification could therefore be a valuable option.  The
SDSS supplies five-band information, along with ready-made tools for the
derivation of photometric redshifts for all host galaxies, and spectral
identification for part of the brighter ones. This makes our approach of using
the SN photometry for classification, assuming that the redshift is known, even
more appropriate. Since no followup is required, such an analysis need not be
done in real time --- in principle, it may even be carried out long after the
project is complete, and all the data are publicly available.

In summary, we have presented a method for the classification of SNe, using
multi-color broadband photometry. We have shown that the type of a SN may be
uniquely determined in some cases. Even when this is not the case, constraints
on the type may still be drawn. In particular, SNe~Ia may be distinguished from
core-collapse SNe during long periods in their evolution, for most of the
redshifts and filter combinations studied. Finally, we have demonstrated the
application of our method to a recently discovered SN from the SDSS, and we
have shown how this technique may become a valuable tool for the analysis of
the large SN samples expected to emerge from this and other programs that are
already active or will begin soon. This work proves that spectroscopic followup
is not always a prerequisite for SNe to be a valuable scientific resource, and
lays the foundations for the exploitation of large SN samples for which such
followup may not be possible.

\acknowledgments

We thank the many people in A.V.F.'s group at the University of California,
Berkeley (especially A. J. Barth, R. Chornock, L. C. Ho, and J. C. Shields)
who, over the years, helped obtain and calibrate the spectra that constitute
much of the dataset used in this study. We are grateful to A. Fassia, M. Hamuy,
and Y. Qiu for supplying us with digital copies of their spectra of SN 1998S,
1999ee, and 1996cb, respectively. B. Leibundgut generously provided help with
SN light curves, and we thank D. Branch, E. O. Ofek, G. Richards, and
O. Shemmer for useful suggestions.  We acknowledge the assistance of the staffs
of various observatories (especially Lick) where the data were taken.  This
research has made use of the NASA/IPAC Extragalactic Database (NED), which is
operated by the Jet Propulsion Laboratory, California Institute of Technology,
under contract with the National Aeronautics and Space Administration.
D. M. acknowledges support by the Israel Science Foundation --- the Jack Adler
Foundation for Space Research, Grant 63/01-1.  A.V.F. is grateful for the
financial support of NSF grants AST--9417213 and AST--9987438, as well as of
the Guggenheim Foundation.

\clearpage

\begin{figure*}
\plotone{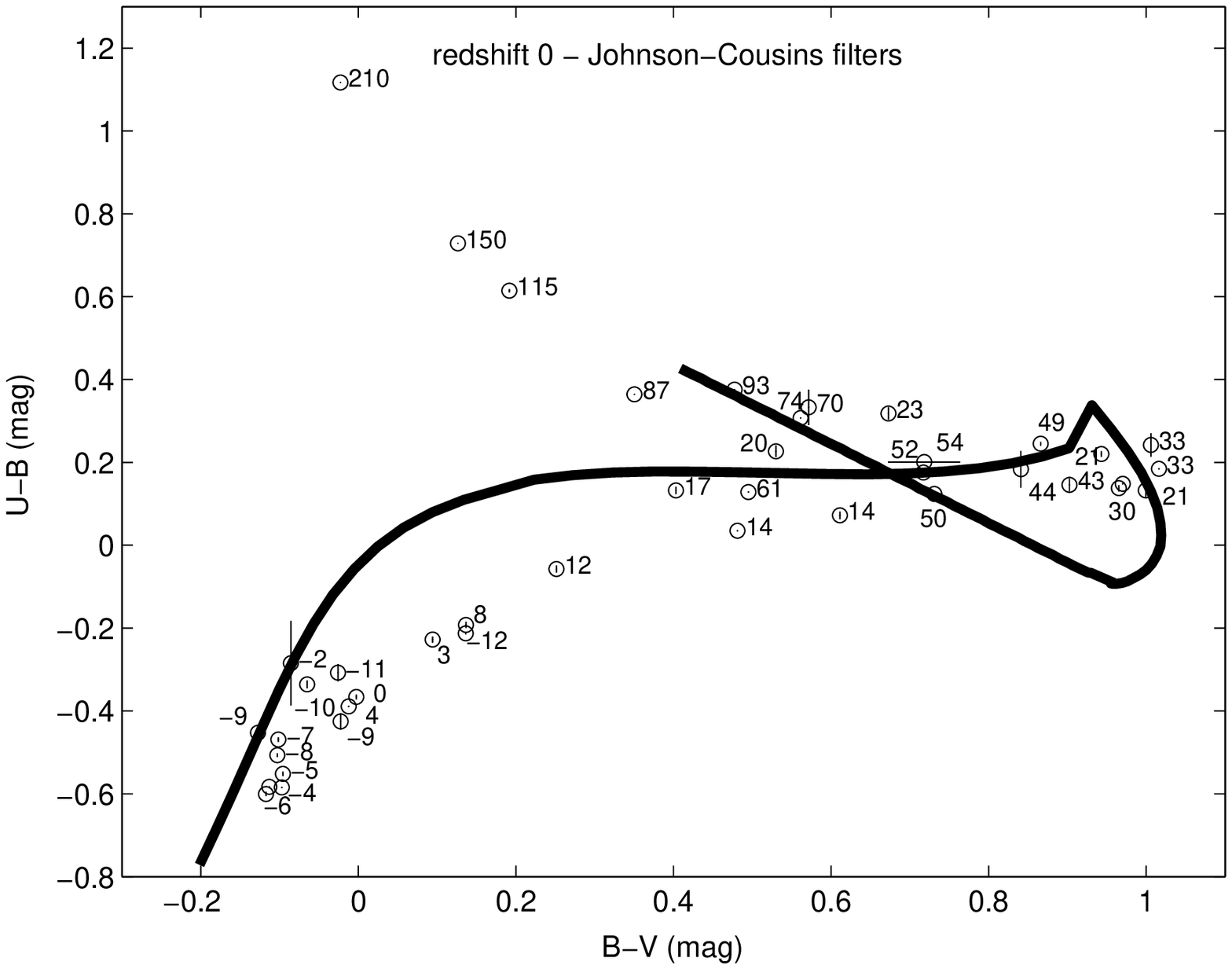}
\caption{Comparison between the synthetic colors (open circles, error bars
calculated as explained in $\S~3.1$) of SNe~Ia, and $UBV$ photometry (bold
line) from Leibundgut (1988). The marked ages (here and in the following
figures) are in days relative to $B$-band maximum. For clarity, we have
sometimes omitted the age labels of data points with similar ages and colors.
Many of the early-time (up to 2 weeks past maximum brightness) synthetic
colors were derived from spectra of SN 1994D, which was somewhat bluer 
at those epochs than average SNe~Ia; thus, from this plot we can estimate the
expected width of color-color paths for normal SNe~Ia to be 
$\sim 0.2$--0.3 mag.}
\end{figure*}

\begin{figure*}
\plotone{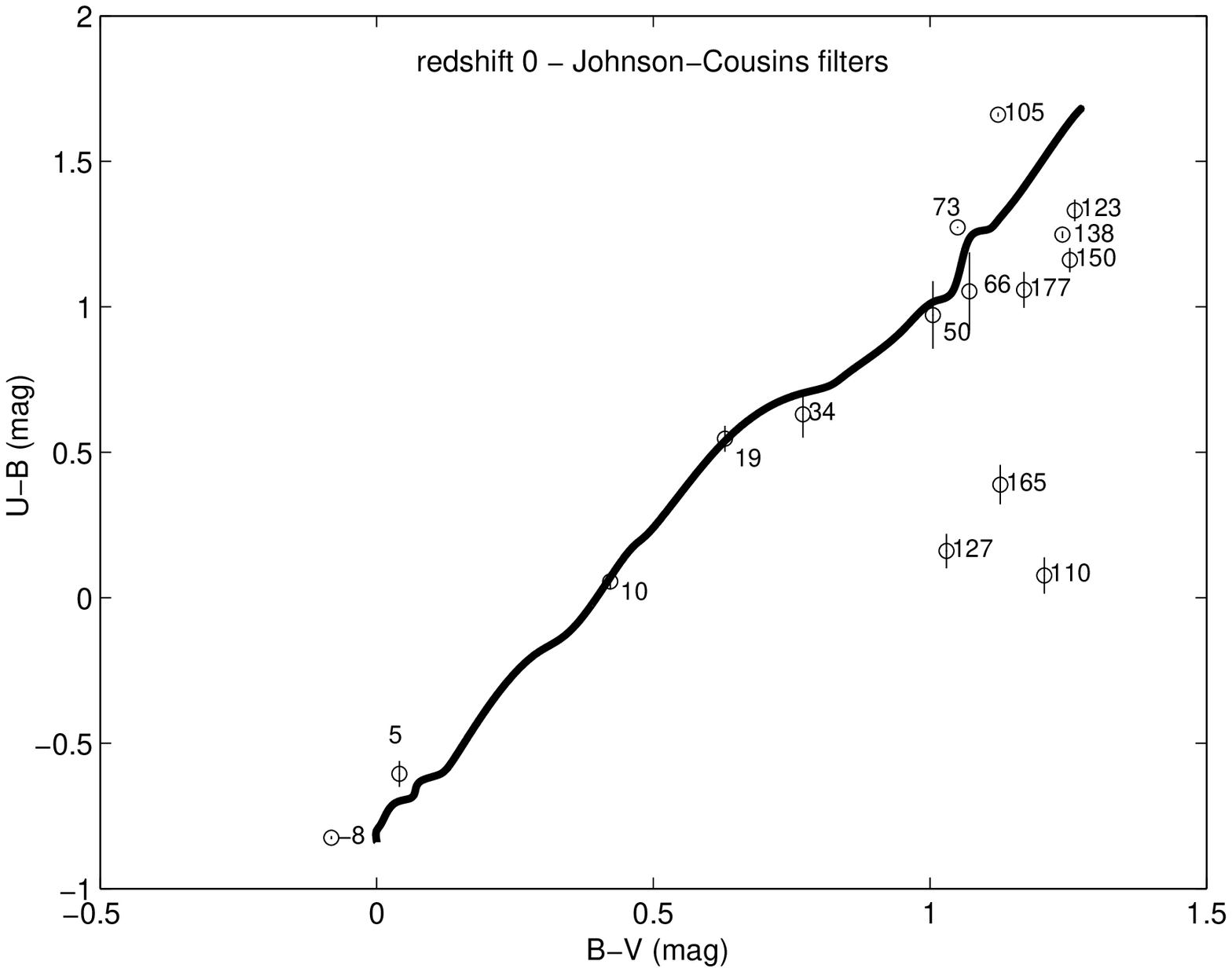}
\caption{Comparison of the synthetic colors (open circles, error bars
calculated as above) for SNe~II-P, with $UBV$ photometry of SN 1999em from
Leonard et al. (2002a) for days $-1$ to 73 relative to $B$-band peak (bold
line). Note that the three discordant points toward the lower-right corner are
derived from spectra of SN 2001X (see $\S~3.1$ for discussion).}
\end{figure*}

\begin{figure*}
\epsscale{0.8}
\plotone{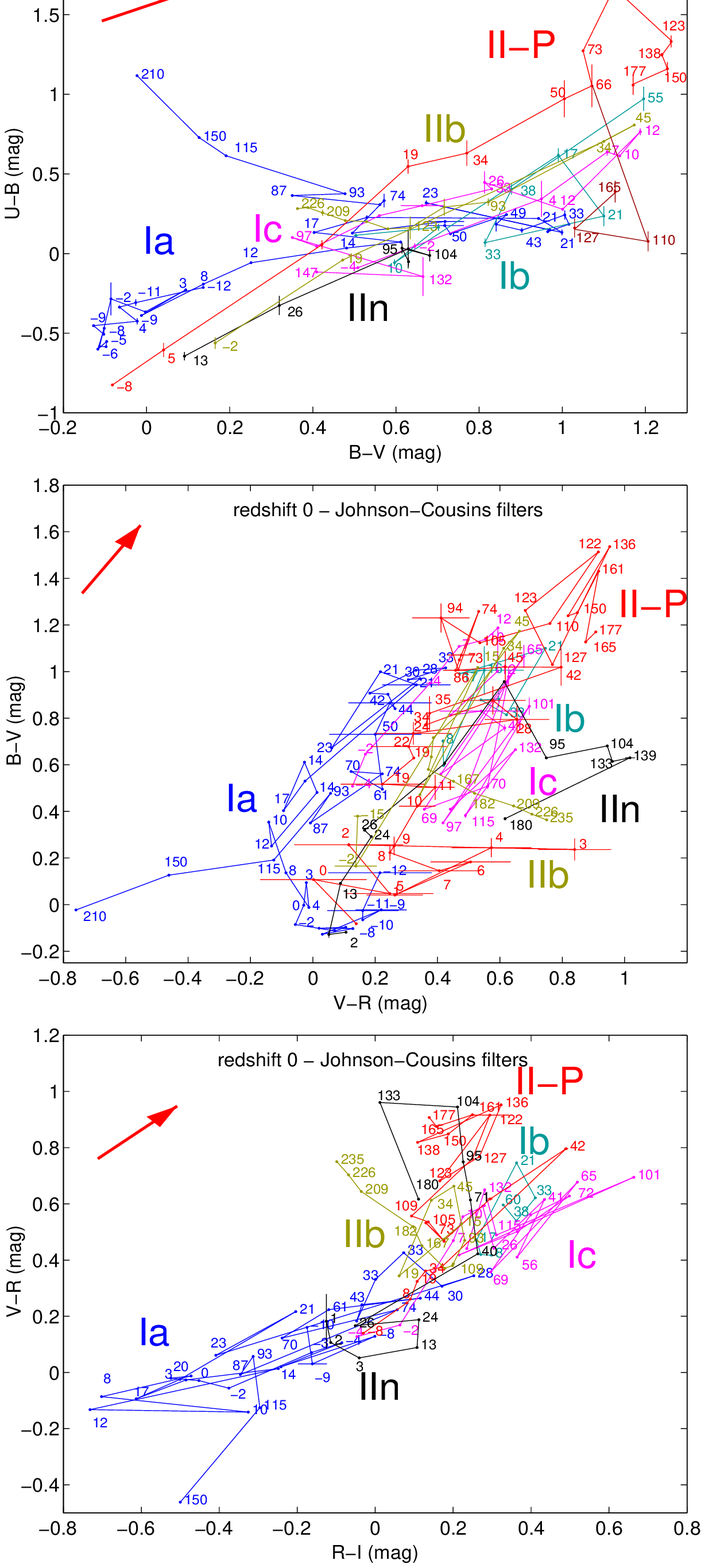}
\caption{$U-B$ vs. $B-V$ (top), $B-V$ vs. $V-R$ (middle), and $V-R$ vs. $R-I$
(bottom) for SNe of all types at $z = 0$.  In this and all subsequent figures,
the time evolution of each SN type is illustrated by linearly connecting, in
temporal order, the locations occupied by such events.  Note that we depict
both branches of late-time evolution of type II-P SNe; see $\S~3.1$.  The arrow
shows the reddening effect of $A_{V} = 1$ mag (in restframe $V$) extinction by
dust in the host galaxy.}
\end{figure*}

\begin{figure*}
\plotone{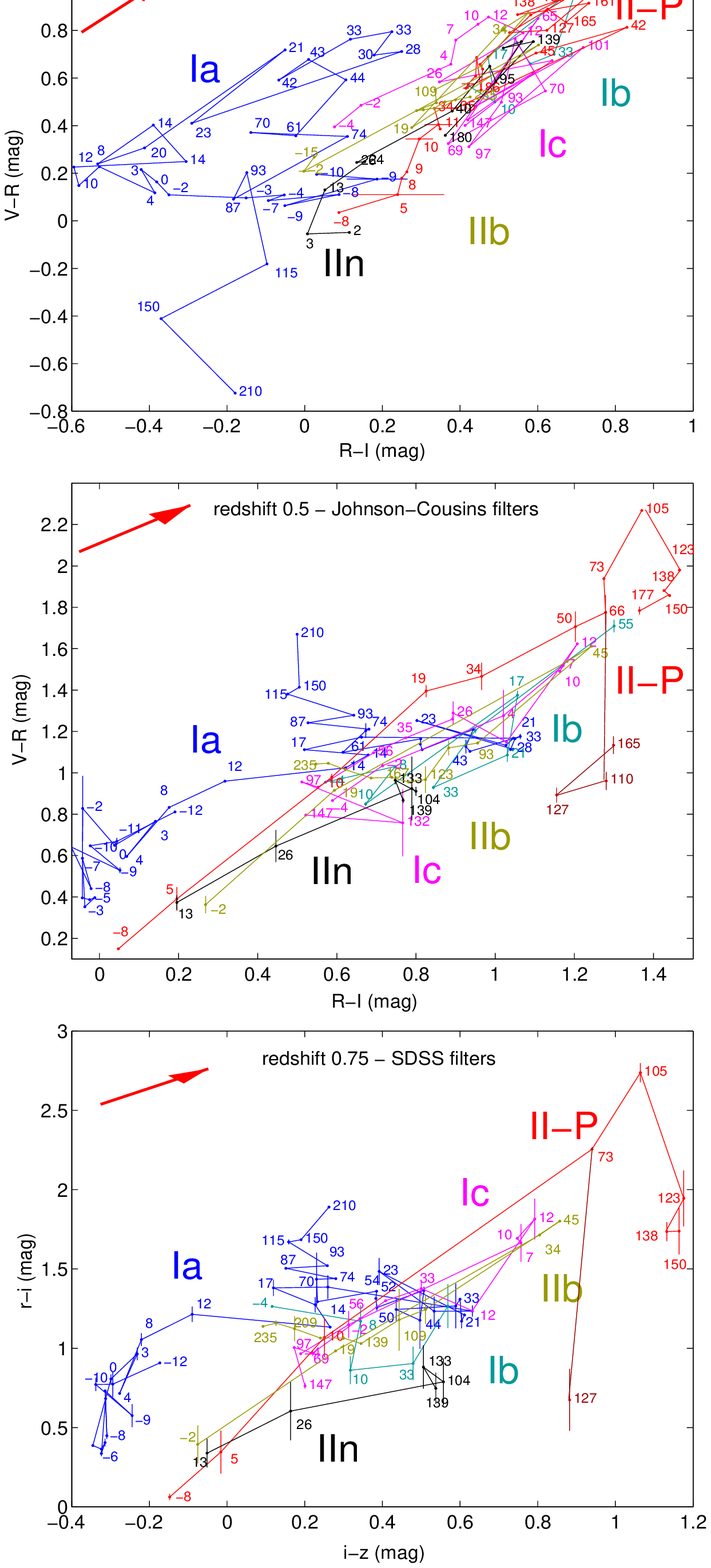}
\caption{$V-R$ vs. $R-I$ for SNe of all types at $z=0.1$ (top) and $z=0.5$
(middle). At bottom, $r-i$ vs. $i-z$ for SNe of all types at $z=0.75$.}
\end{figure*}

\begin{figure*}
\plotone{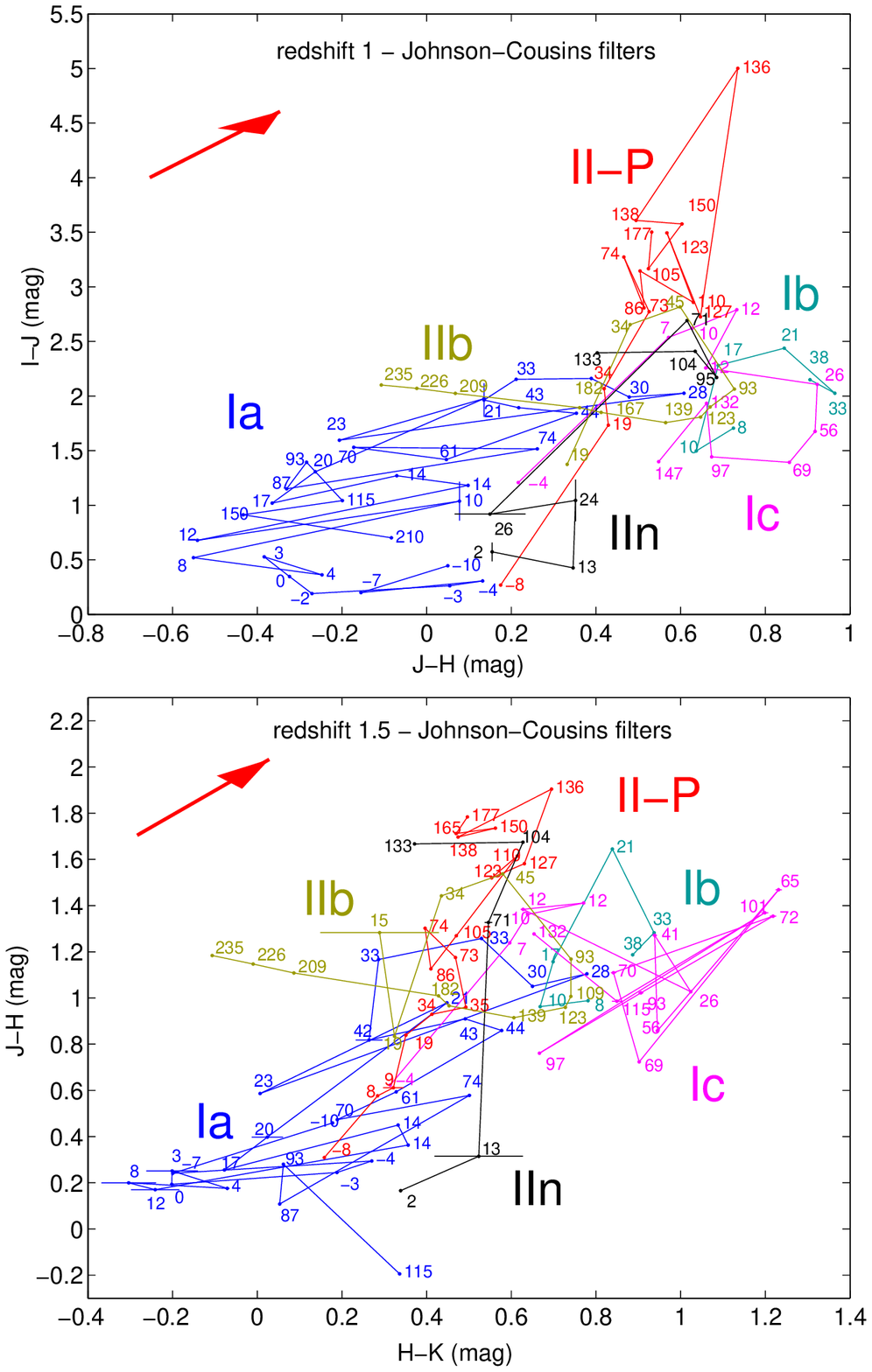}
\caption{$I-J$ vs. $J-H$ for SNe of all types at $z=1$ (top), and
$J-H$ vs. $H-K$ for SNe of all types at $z=1.5$ (bottom).} 
\end{figure*}

\begin{figure*}
\epsscale{0.8}
\plotone{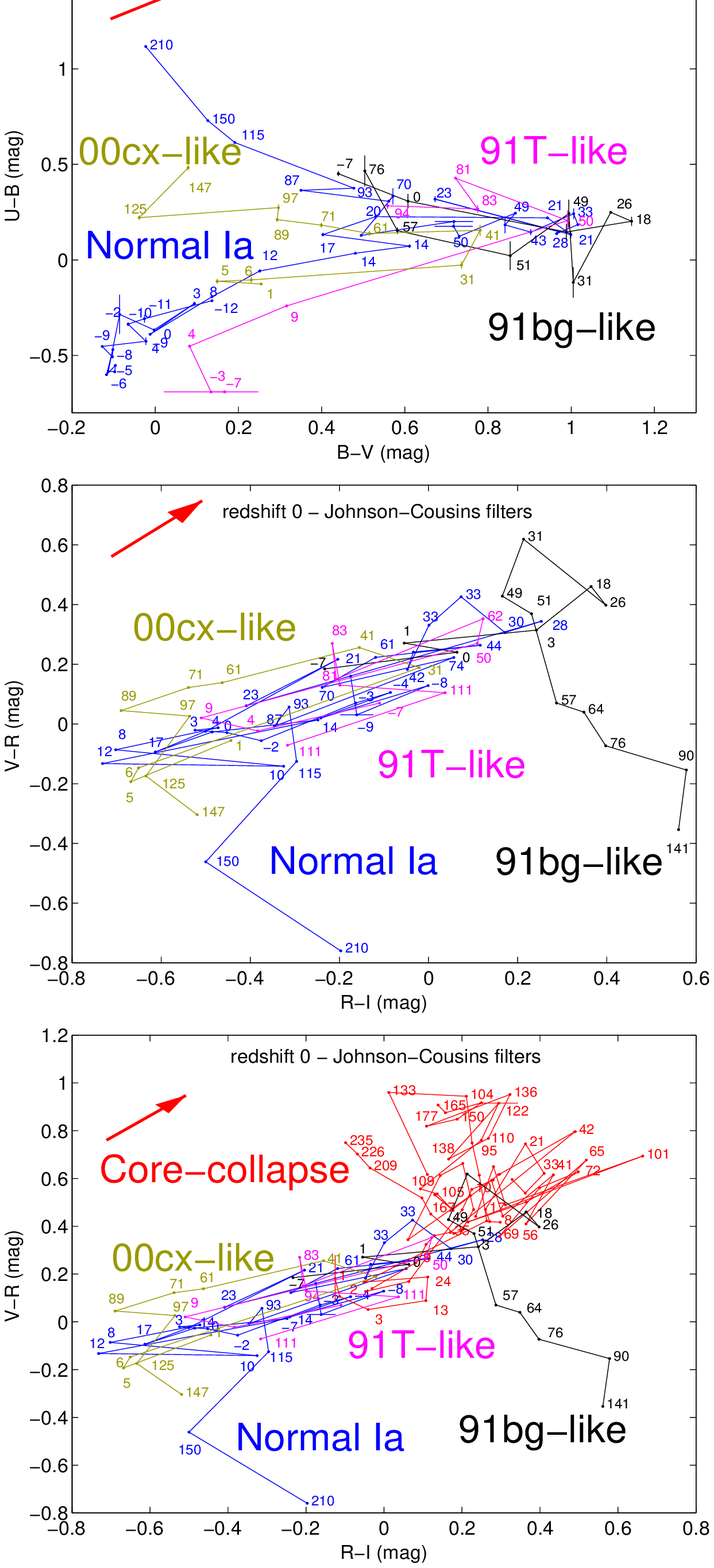}
\caption{$U-B$ vs. $B-V$ (top) and $V-R$ vs. $R-I$ (middle) for normal 
(blue), SN 1991bg-like (black),
SN 1991T-like (magenta), and SN 2000cx-like (green) SNe~Ia at $z=0$.
Bottom: same as top panel, but including
core-collapse SNe of all types (red).}
\end{figure*}

\begin{figure*}
\plotone{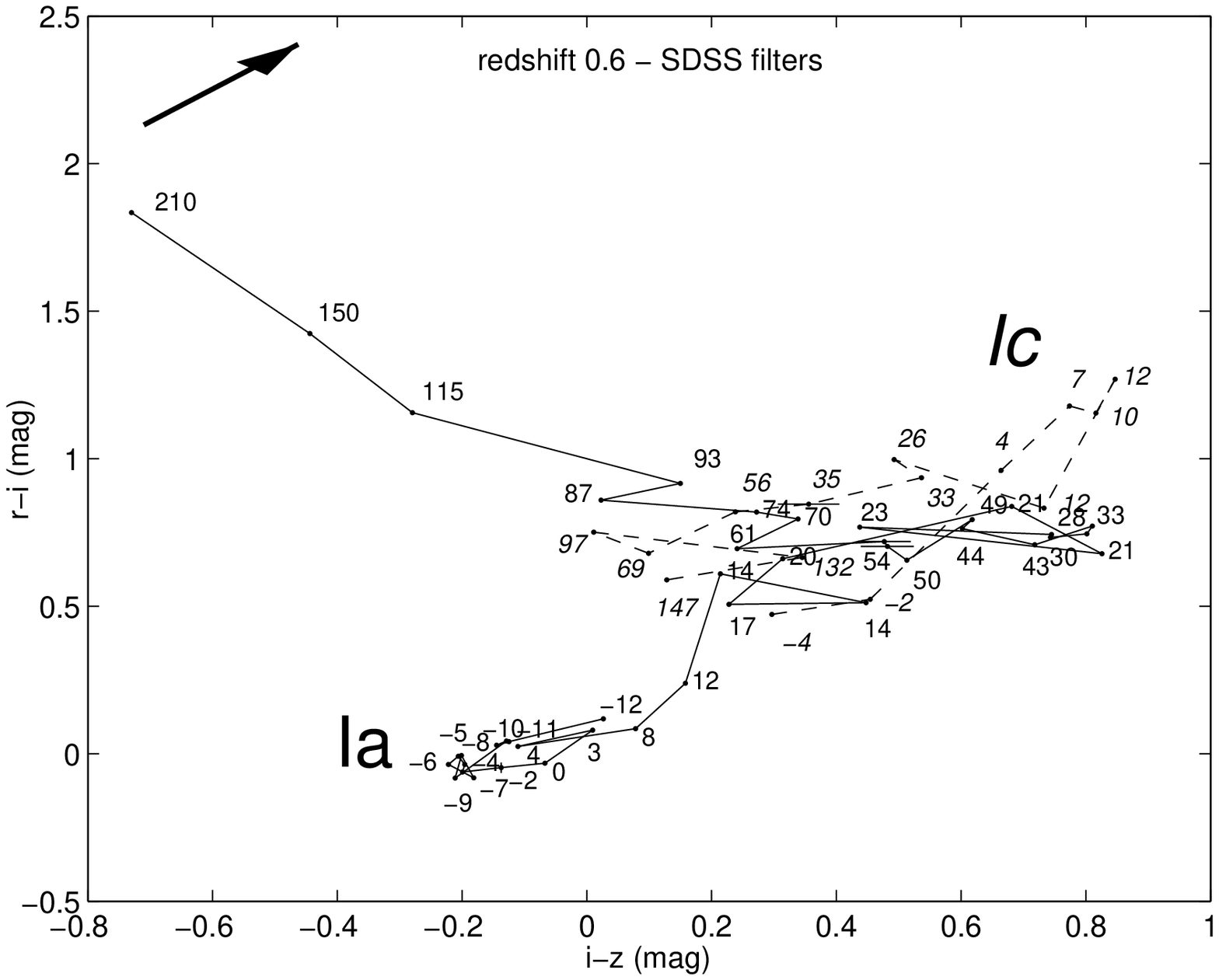}
\caption{$r-i$ vs. $i-z$ for SNe~Ia (black solid lines) and SNe~Ic (grey dashed
lines, italic age labels) at $z=0.6$.}
\end{figure*}

\begin{figure*}
\plotone{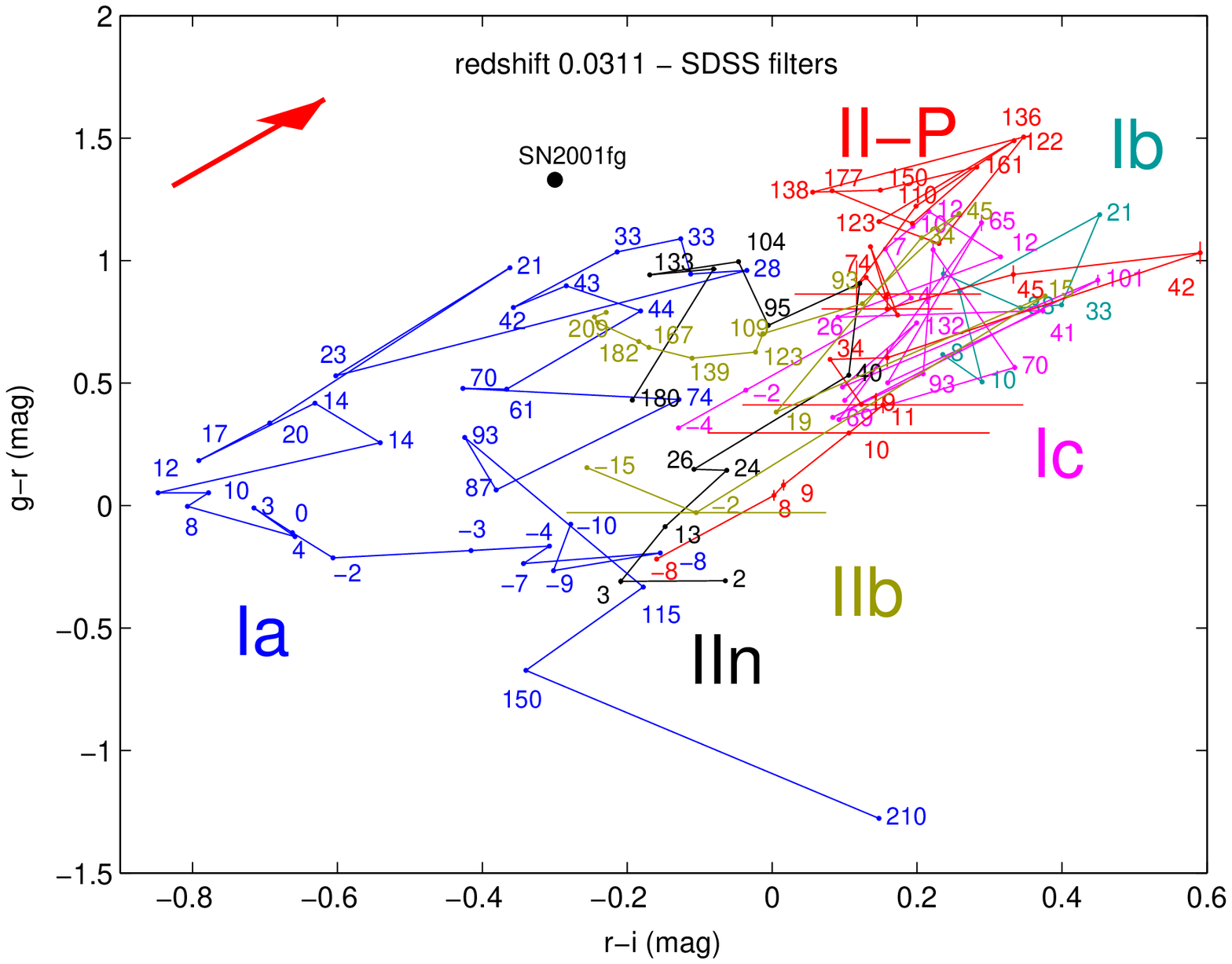}
\caption{$g-r$ vs. $r-i$ for SNe of all types at $z = 0.0311$. The observed
colors of SN 2001fg are indicated by the filled black circle. The colors
suggest a SN~Ia about 1 month past maximum brightness, as confirmed by spectra
of this event.} 
\end{figure*}

\begin{figure*}
\plotone{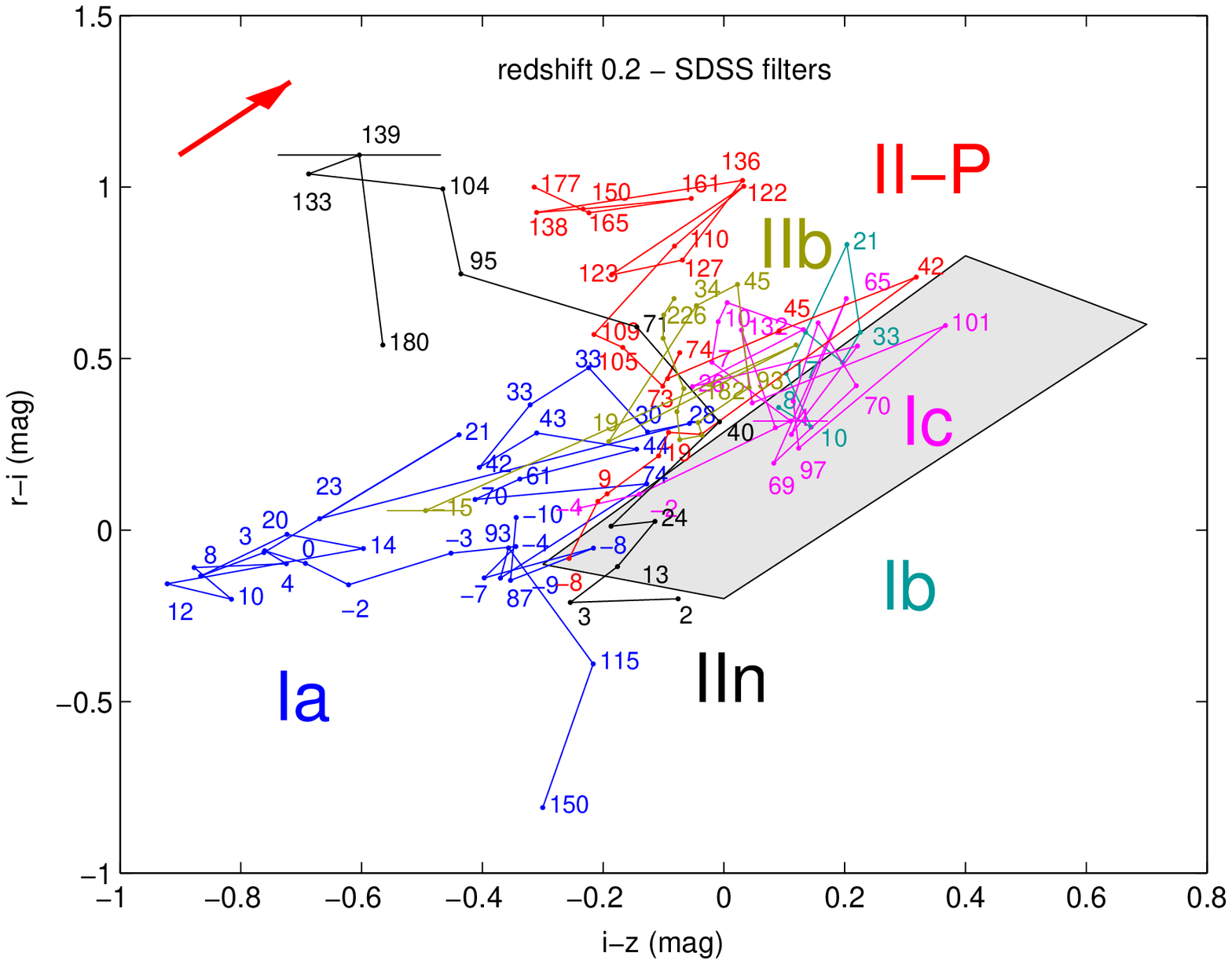}
\caption{$r-i$ vs. $i-z$ for SNe of all types at $z=0.2$. The gray
polygon includes the colors of almost all stars, cataclysmic
variables, Cepheids, asteroids, and quasars from Krisciunas et al. (1998).} 
\end{figure*}





\clearpage
\begin{deluxetable}{clccc}
\tabletypesize{\scriptsize}
\tablecaption{SN Spectral Database \label{database table}}
\tablewidth{0pt}
\tablehead{
\colhead{Type} & \colhead{SNe} & \colhead{Epochs} & \colhead{Redshift}&
\colhead{References\tablenotemark{a}} 
}
\startdata
Ia   & 1994D  & 22 & 0.0015 & 1,9 \\
     & 1987L  & 2  & 0.0074 & 1\\
     & 1995D  & 4  & 0.0066 & 9\\
     & 1999dk & 5  & 0.0150 & 9\\
     & 1999ee & 12 & 0.0114 & 10\\
Ib\tablenotemark{b}   & 1984L  & 12 & 0.0051& 1,9 \\
     & 1991ar & 1 & 0.0152 & 2 \\
     & 1998dt & 2 & 0.0150 & 2 \\
     & 1999di & 1 & 0.0164 & 2 \\
     & 1999dn & 3 & 0.0093 & 2 \\
Ic   & 1994I  & 14 & 0.0015 & 1,3 \\
     & 1990U  & 8 & 0.0079 & 2 \\
     & 1990B  & 4 & 0.0075 & 2 \\
II-P & 1999em & 27 & 0.0024 & 4\\
     & 1992H  & 13 & 0.0060 & 1,9\\
     & 2001X  & 12 & 0.0049 & 6,9\\
IIn  & 1998S  & 13 & 0.0030 & 5,7\\
     & 1994Y  &  1 & 0.0080 & 1 \\
     & 1994ak &  1 & 0.0085 & 1 \\
IIb  & 1993J  & 12 & 0 & 1 \\
     & 1996cb &  3 & 0.0024 & 8 \\
\tableline
Total & & 172 &&\\
\enddata

\tablenotetext{a}{
(1) Filippenko 1997; (2) Matheson et al. 2001; (3) Filippenko et al. 1995b;
(4) Leonard et al. 2002a; (5) Leonard et al. 2000; (6) Gal-Yam \& Shemmer
2001; (7) Fassia et al. 2001; (8) Qiu et al. 1999;
(9) Unpublished spectra by Filippenko and collaborators, obtained
and reduced as those presented in (1)--(5); (10) Hamuy et al. 2002.\\}
\tablenotetext{b}{In the case of the SNe~Ib 1998dt, 1999dn, 1999di, and
1991ar the temporal order used is from Matheson et al. (2001) derived from
the relative depth of helium lines in the spectra.\\}

\end{deluxetable}
   
\clearpage

\begin{deluxetable}{clccc}
\tabletypesize{\scriptsize}
\tablecaption{Peculiar Type Ia SNe \label {pecIa table}}
\tablewidth{0pt}
\tablehead{
\colhead{Type} & \colhead{SNe} & \colhead{Epochs} &
\colhead{Redshift} & \colhead{References\tablenotemark{a}} 
}
\startdata
1991T-like & 1991T  & 12 & 0.0058 & 1,5 \\
     & 1998es & 7 & 0.0106 & 5 \\
1991bg-like & 1991bg & 5  & 0.0035 & 2,5 \\
     & 1998bp & 2 & 0.0104 & 5 \\
     & 1998de   & 3 & 0.0166 & 5 \\
     & 1999da   & 4 & 0.0127 & 5 \\
2000cx-like & 2000cx & 23 & 0.0079 & 3,4 \\
\tableline
Total & & 56 &&\\

\enddata
\tablenotetext{a}{
(1) Filippenko et al. 1992a; (2) Filippenko et al. 1992b;  (3) Li et al.
2001b; (4) Gal-Yam \& Shemmer 2001;
(5) Unpublished spectra by Filippenko and collaborators, obtained 
and reduced as those presented in (1)--(3).\\}
\end{deluxetable}

\end{document}